\documentclass[aps,prc,twocolumn,floatfix,superscriptaddress]{revtex4}
\usepackage{epsfig}
\usepackage{amsmath}
\usepackage{color}
\usepackage[utf8]{inputenc}
\usepackage{graphicx}

\definecolor{blue}{rgb}{0.05, 0.05, 0.5}

\begin{document}

\title{Landau Pomeranchuk Midgal Effect and Charm Production in $pp$ Collisions at Large Hadron Collider Energies using the Parton Cascade Model}
\author{Dinesh K. Srivastava}
\email{dinesh@vecc.gov.in}
\affiliation{Variable Energy Cyclotron Centre, HBNI, 1/AF, Bidhan Nagar, Kolkata 700064, India~\footnote{Present and Permanent Address}}
\affiliation{Institut für Theoretische Physik, Johann Wolfgang Goethe-Universität, Max-von-Laue-Str. 1, D-60438 Frankfurt am Main, Germany}
\affiliation{ExtreMe Matter Institute EMMI, GSI Helmholtzzentrum für Schwerionenforschung, Planckstrasse 1, 64291 Darmstadt, Germany}
\author{Rupa Chatterjee}
\email{rupa@vecc.gov.in}
\affiliation{Variable Energy Cyclotron Centre, HBNI, 1/AF, Bidhan Nagar, Kolkata 700064, India}
\author{Steffen A. Bass}
\email{bass@phy.duke.edu}
\affiliation{Duke University, Department of Physics, 139 Science Drive,
Box 90305, Durham NC 27708, U. S. A.}

\begin{abstract}
We study the impact of the Landau Pomeranchuk Midgal (LPM) effect on the dynamics of parton interactions in proton proton collisions at the Large Hadron Collider energies. For our investigation we utilize  a  microscopic kinetic theory based on the Boltzmann equation. 
The calculation
traces the space-time evolution of the cascading partons interacting via semihard pQCD scatterings and fragmentations. We focus on the impact of the LPM effect on the production  of charm quarks, since their production is exclusively governed by processes well described in our kinetic theory. The LPM effect is found to become more prominent as the collision energy rises 
and at central rapidities and may significantly affect the model's predicted charm distributions at low momenta.
\end{abstract}


\maketitle

\maketitle
\section{Introduction}
Studies of relativistic collisions of heavy nuclei underway at the Relativistic Heavy
Ion Collider at Brookhaven and the Large Hadron Collider at CERN have provided
ample evidence for a deconfining transition of strongly interacting matter into a (strongly 
coupled) Quark Gluon Plasma (QGP) expected from lattice QCD calculations 
(see e.g., Refs.~\cite{Ratti:2016lrh,Ratti:2017qgq,Ratti:2018ksb} and references therein).
 These studies, both on the
theoretical and the experimental fronts, have now reached a high level of sophistication
and the quantitative determination of QGP properties \cite{Schenke:2010rr,Gale:2012rq,Shen:2014vra,Bernhard:2016tnd}  is now in progress. Very often the results for heavy-ion collisions
are compared with those for proton proton collisions at the same center of mass energy ($\sqrt{s_{NN}}$) in order to arrive at some of
these conclusions, with the rationale that no QGP is likely to be formed in $pp$ collisions. This
simple expectation is now under strain as more and more indications of formation of an interacting 
system, emerge in $pp$ collisions, especially for events having a large particle 
multiplicity (see e.g., Refs.~\cite{Khachatryan:2016txc,ALICE:2017jyt}).

Is an interacting system formed in $pp$ collisions? 
Recently we have explored this question within Parton Cascade Model (PCM) \cite{Srivastava:2018dye}.  
The PCM is a transport model based on the relativistic Boltzmann equation for the time evolution of the parton density in phase-space due to semi-hard perturbative QCD interactions including scattering and radiations \cite{Geiger:1991nj, Bass:2002fh} within a leading logarithmic approximation~\cite{Altarelli:1977zs}. Our study indicated the formation of a medium driven by a substantial amount of multiple parton
interactions, including fragmentation of partons after scattering. These aspects were found
to be more strongly prevalent for collisions at small impact parameters or with large parton multiplicities and at higher incident beam energies. Even though the precise number of collisions and fragmentations are 
dependent on the $p_T^\text{cut-off}$ and $\mu_0$ used to regularize the pQCD cross-sections and the fragmentations respectively, the results are sufficiently general.

Based on these previous findings it is opportune to investigate the importance of quantum coherence effects in parton-parton interactions, such as the Landau Pomeranchuk Midgal (LPM) effect~\cite{Landau:1953gr}. The LPM effect is known to be important for large collision systems with lifetimes of multiple fm/c, but has commonly been neglected in the microscopic study of the proton-proton system, due to its small size and short lifetime. 

Here we focus on the investigation of the LPM effect on charm quark production in proton proton collisions. Charm production is particularly well suited in this context, since it only occurs via hard processes calculable in pQCD and charm is conserved throughout the reaction. The PCM was recently extended to treat the production and medium-evolution of heavy quarks~\cite{Srivastava:2017bcm}.

Consider a parton traversing a cloud of quarks and gluons and undergoing multiple scatterings. If the separation between consecutive scatterings suffered by the parton is sufficiently large so that the radiations off these collision centers can be treated as an incoherent sum of radiation spectra resulting from individual scatterings, we reach what is known as the Bethe-Heitler 
limit~\cite{Bethe:1934za}. If on the other hand, the scattering centers are too closely located to each other, the observed radiation has to be evaluated within what is known as the factorization limit, and is a product of a single scattering spectrum from the sum of the  individual small momentum transfers from all the individual scatterings. 

The LPM effect~\cite{Landau:1953gr}  describes the results between these two limiting regimes,
by accounting for the suppression of the radiation relative to the Bethe-Heitler limit, when the formation time of the radiated gluon is large compared to the mean free path and thus destructive interference 
between the radiated spectra becomes important. The dynamics of LPM effect on the production of light 
partons ($u$, $d$, $s$, and $g$) and photons in collision of gold nuclei at RHIC energy, within the PCM, was discussed 
earlier~\cite{Renk:2005yg,Bass:2002vm,Bass:2007hy}. That work also demonstrated that the inclusion of the LPM effect greatly improved the 
agreement of the scaling of multiplicity distributions in $pp$ collisions up to 200 GeV.

\begin{figure}[h]
\centerline{\includegraphics*[width=7.6 cm]{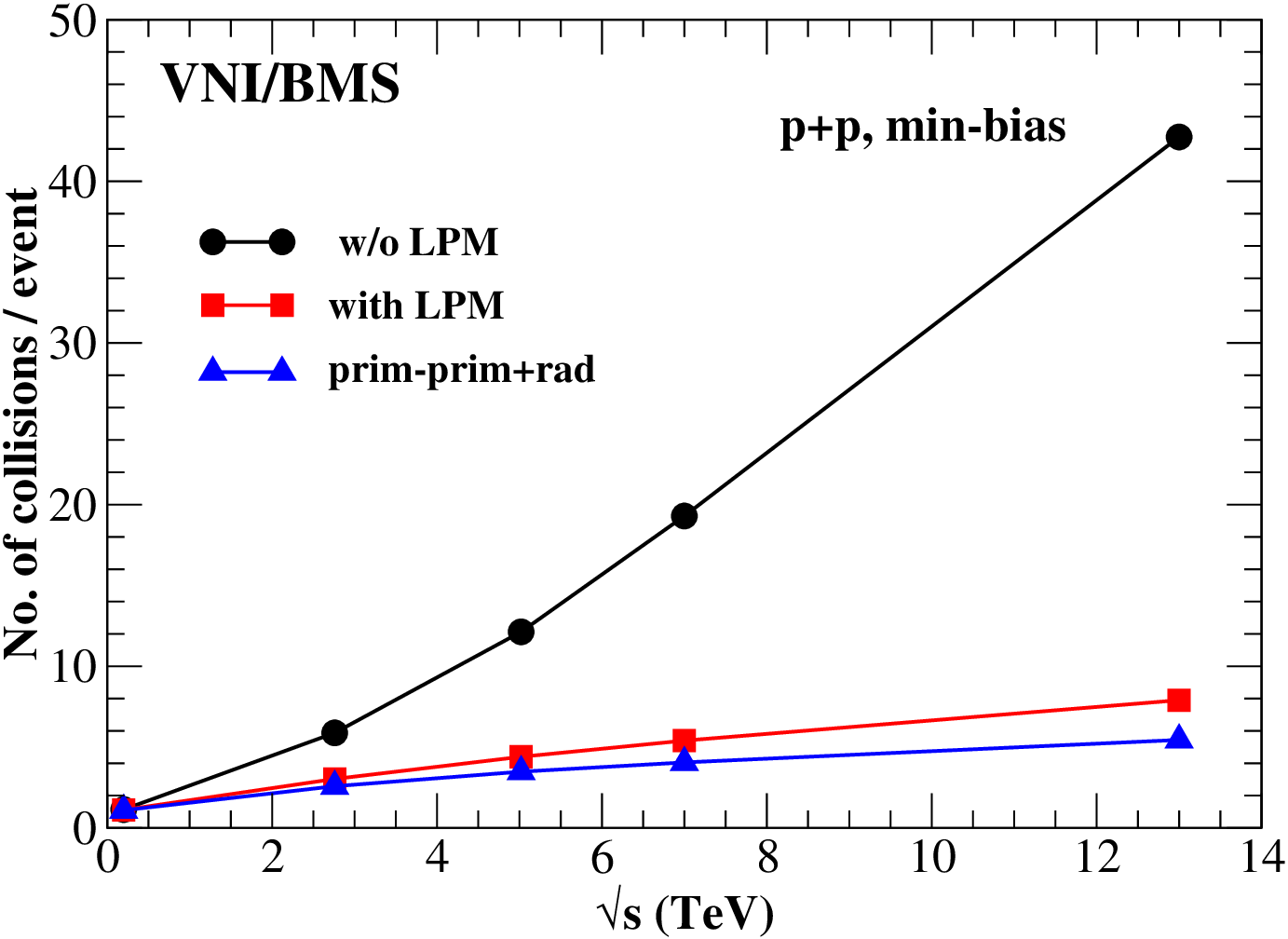}}
\centerline{\includegraphics*[width=7.6 cm]{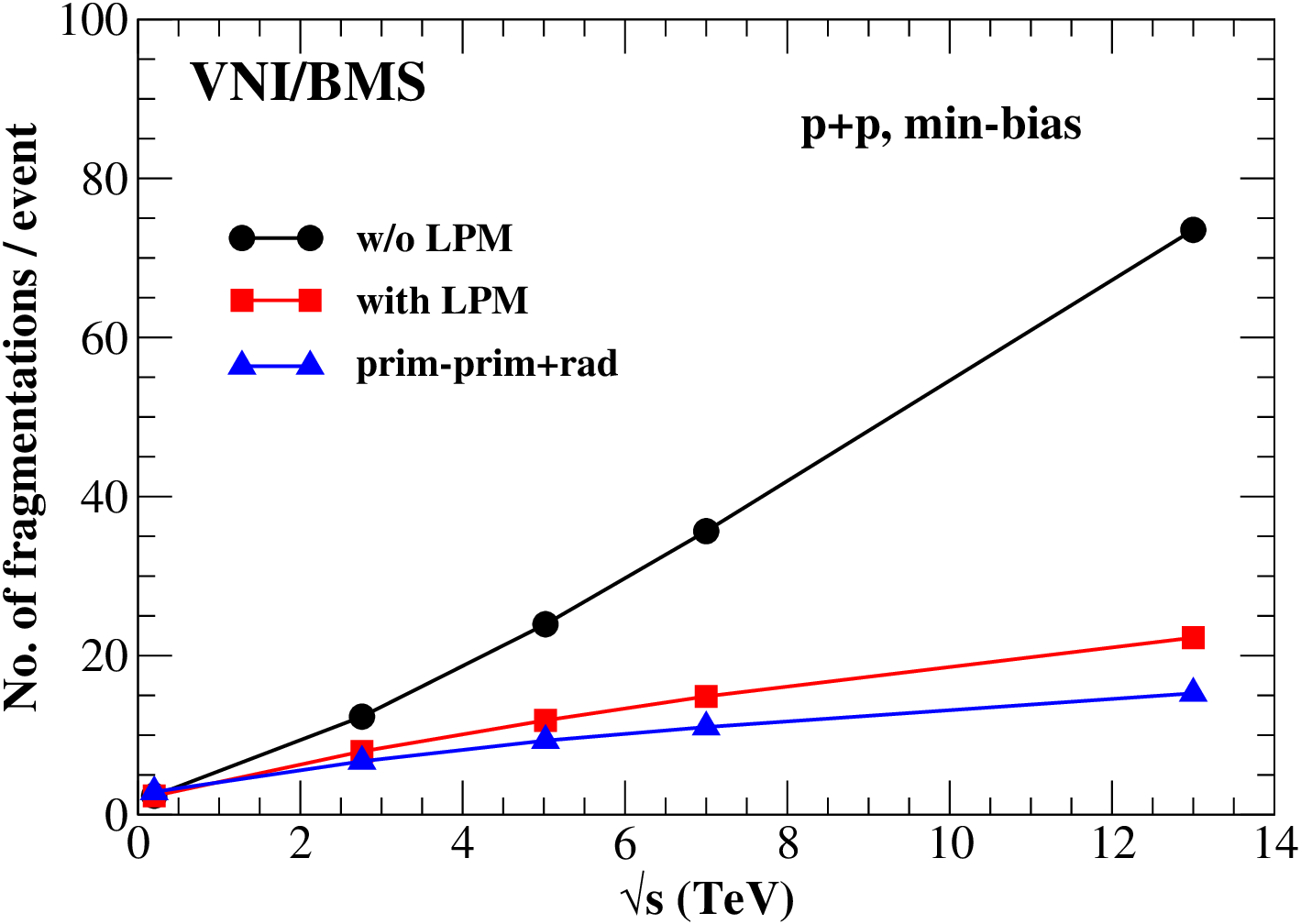}}
\centerline{\includegraphics*[width=7.6 cm]{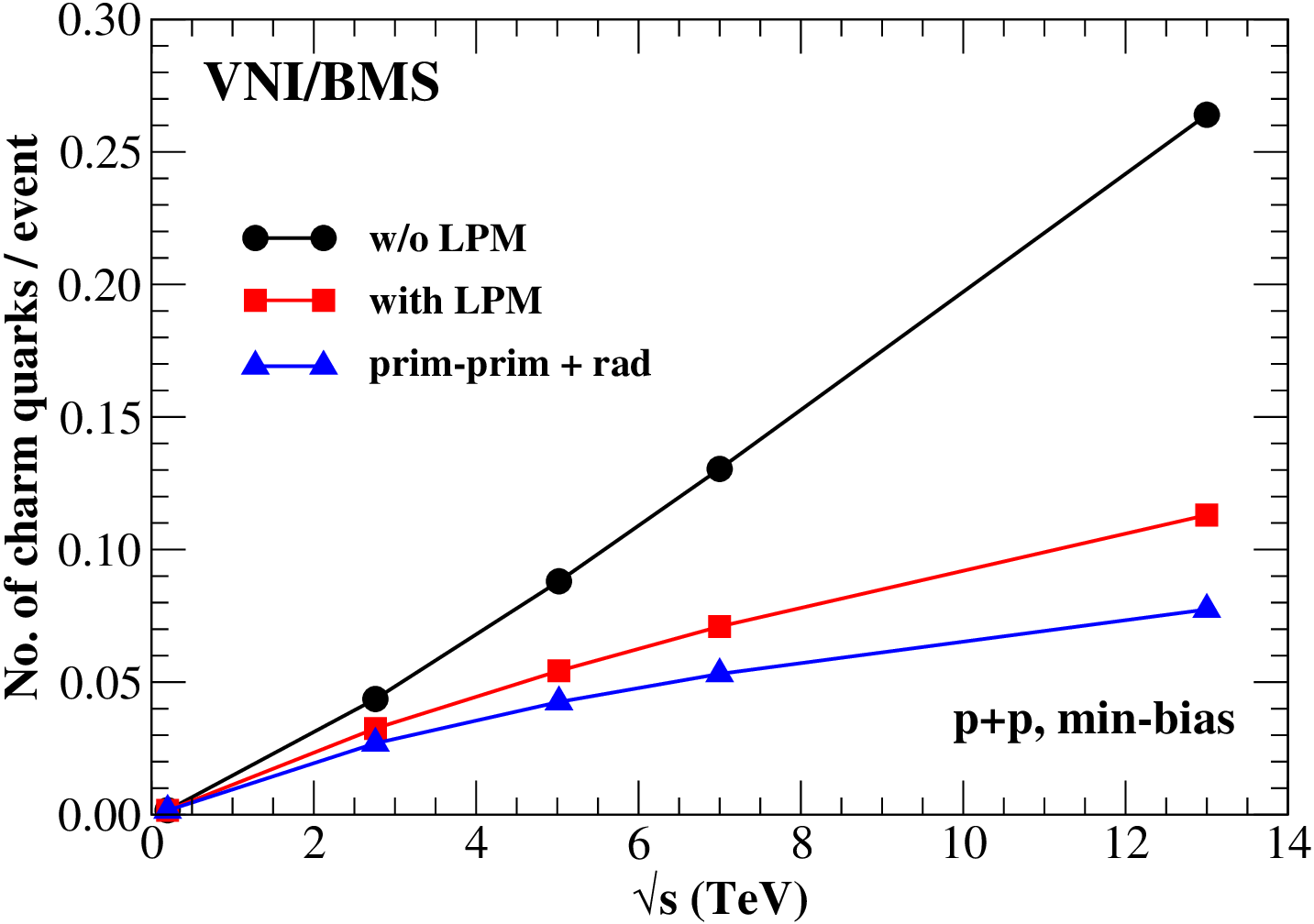}}
\caption{(Color online) Number of collisions (upper panel), number of fragmentations
(middle panel) and number of charm quarks produced per event (lower panel) 
for minimum bias $pp$ interactions as a function of center of mass energy. The 
three calculations involve multiple collisions among partons by neglecting and including the
LPM effect and collisions only among primary partons with radiations off the scattered partons.
}  
\label{min-bias}
\end{figure}

We shall investigate the consequences of the LPM effect on charm production in
 $pp$ collisions at $\sqrt{s_\text{NN}}$ of 0.20, 2.76, 5.02, 7.00, and 13.00 TeV. The results at RHIC
energy (0.20 TeV) are included to clearly bring out the abundance of parton production etc. at
LHC energies.  

There are several reasons for focusing on charm quarks.
As pointed out above, charm quarks can be produced only from semi-hard scattering of gluons and annihilation of a quark-antiquark pair or from a splitting of a gluon which has a large virtuality following a semi-hard scattering.  The corresponding scattering matrix elements are not singular because of the mass of the charm quark
and thus do not need any $p_T^\text{cut-off}$. We do realize, though,  that the
momentum distribution of the charm quarks can be modified by radiation of gluons or by scattering
with other partons, which will be affected by variation of the $p_T^\text{cut-off}$ used for
regularizing the pQCD matrix elements and the $\mu_0$ used for terminating the fragmentations.
The number of charm quarks which are produced is very small and thus the probability that their 
number is depleted by charm-anticharm annihilation is limited.
Finally, there is no production of charm quarks during the hadronic phase. 

We briefly discuss the basic ingredients of the PCM model pertaining to this investigation in the next section, results are given in Section III, and finally we summarize our findings.

\begin{figure}
\centerline{\includegraphics*[width=7.6 cm]{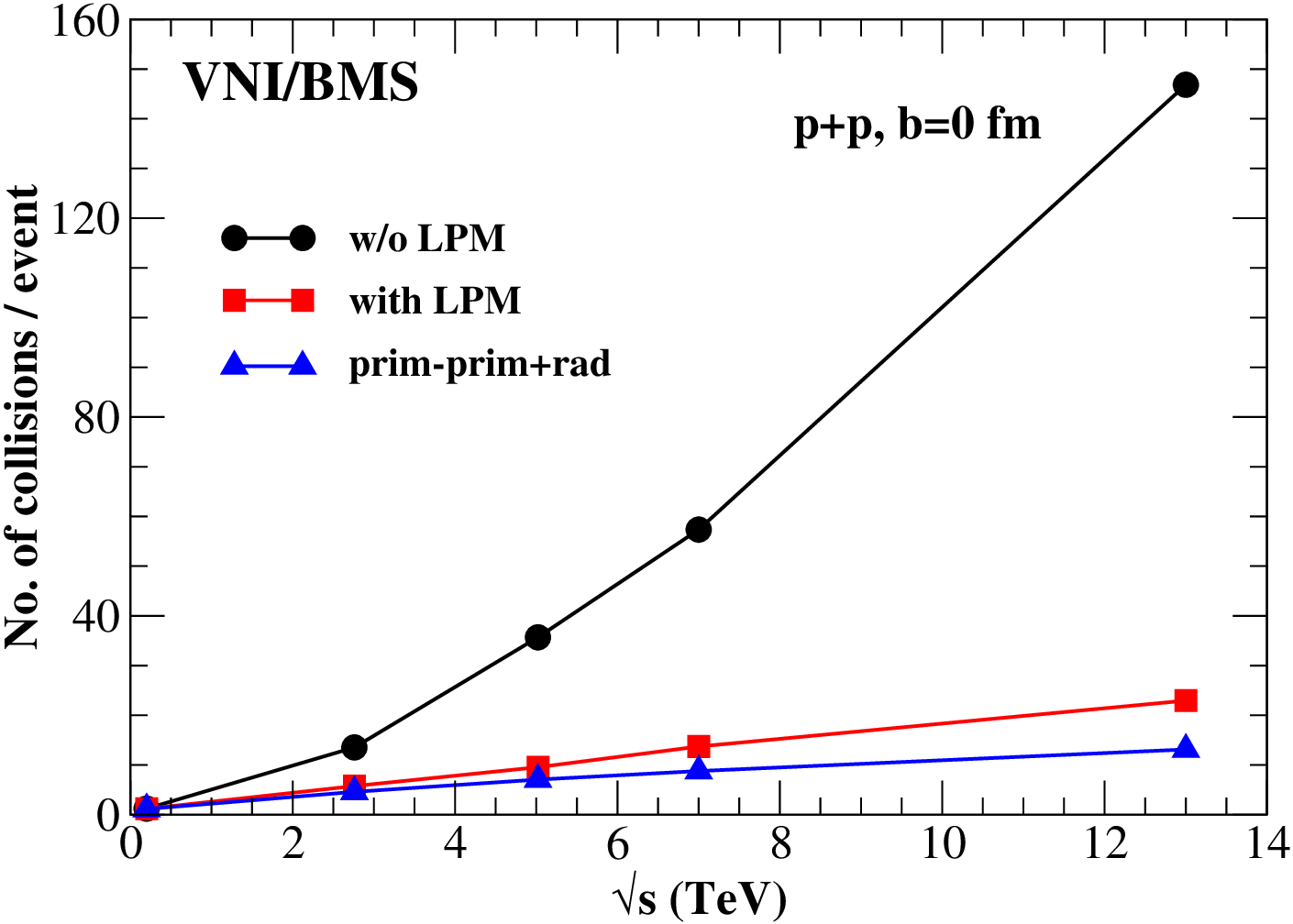}}
\centerline{\includegraphics*[width=7.6 cm]{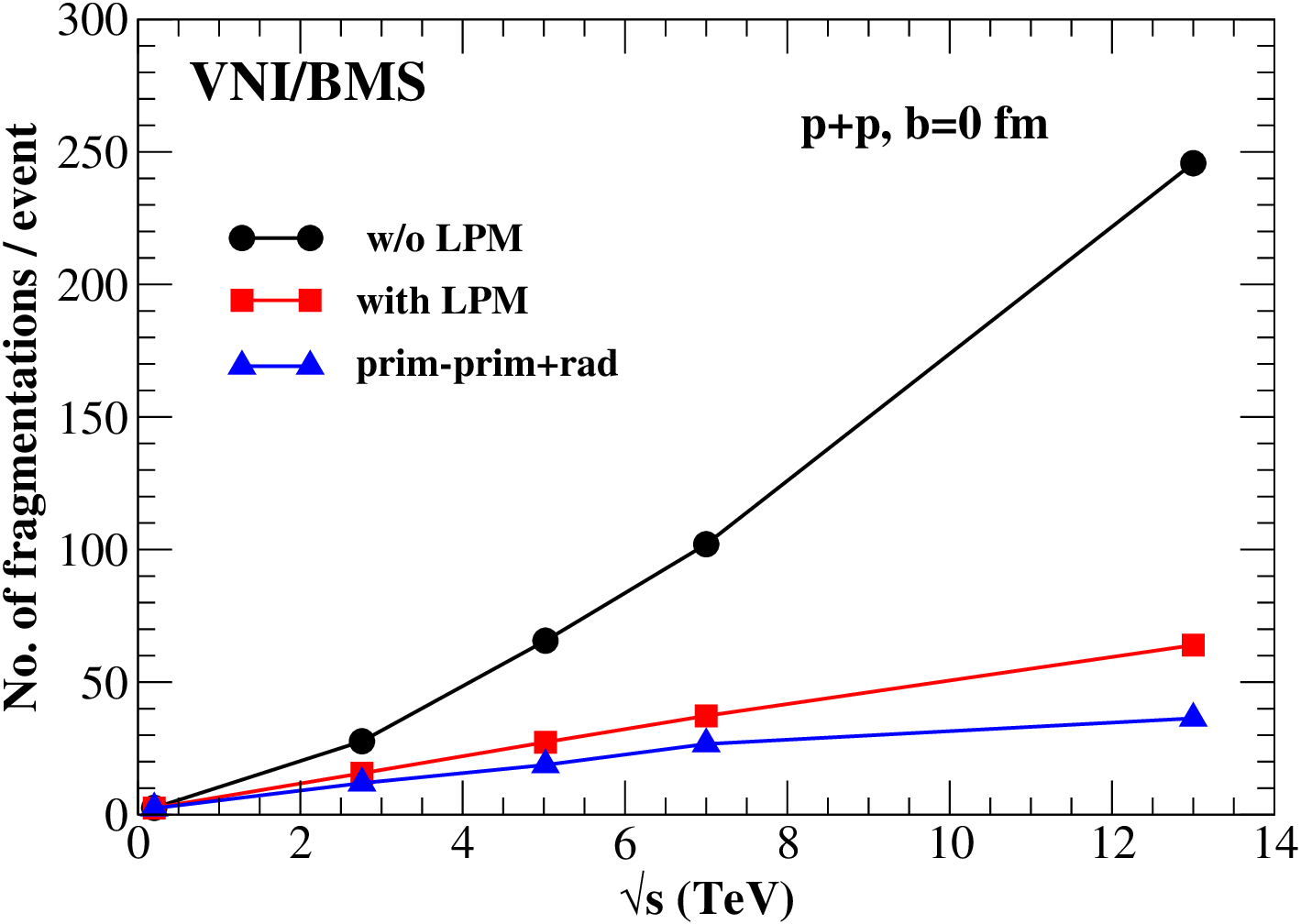}}
\centerline{\includegraphics*[width=7.6 cm]{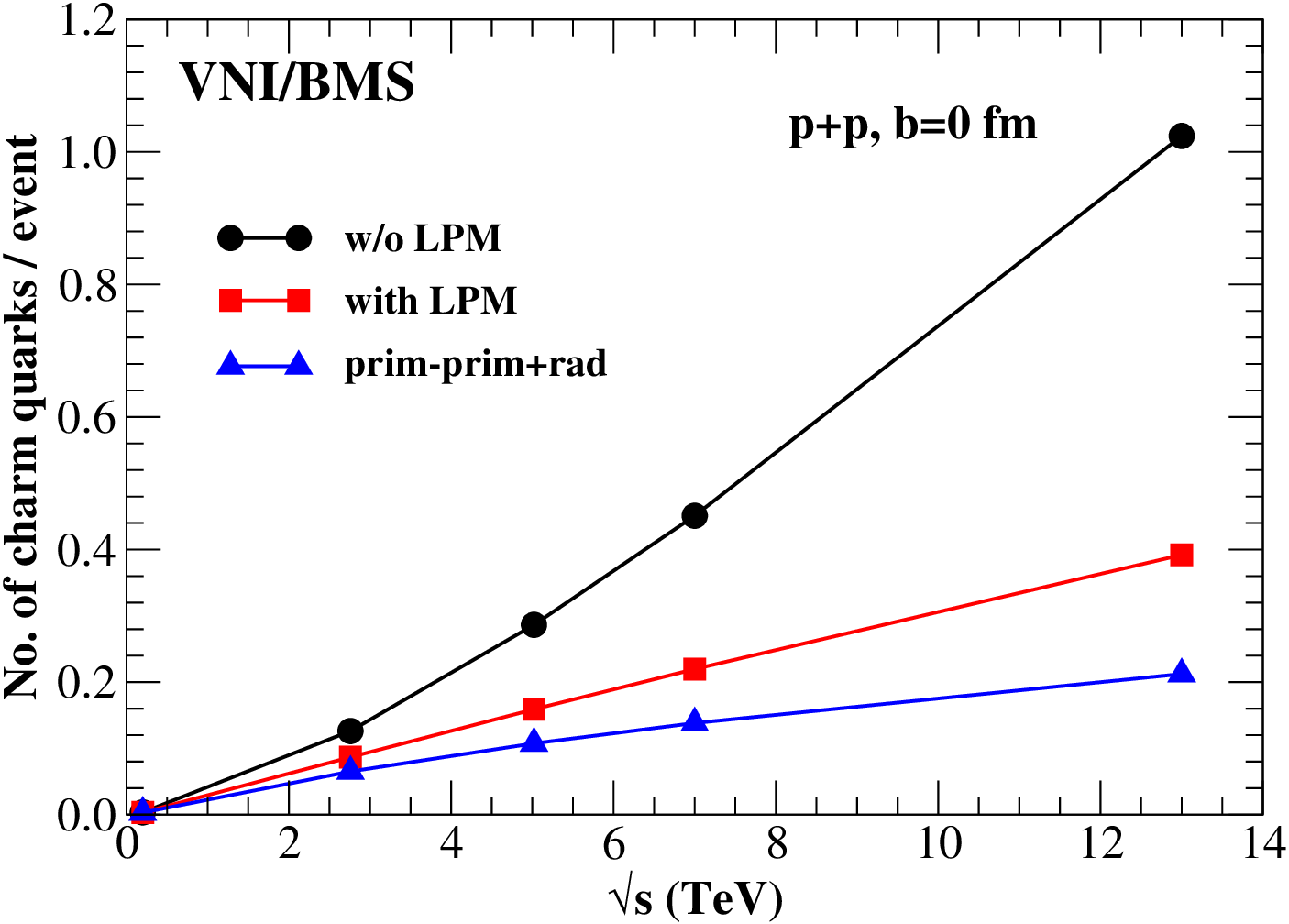}}
\caption{(Color online) Number of collisions (upper panel), number of fragmentations
(middle panel) and number of charm quarks produced per event (lower panel) 
for $pp$ interactions as a function of center of mass energy at impact parameter equal to zero fm. The 
three calculations involve multiple collisions among partons by neglecting and including the
LPM effect and collisions only among primary partons with radiations off the scattered partons.}  
\label{b.eq.0}
\end{figure}

\section{Model Description} 

The details of the parton cascade model, including its  Monte Carlo implementation  {\tt VNI/BMS}, have been discussed in significant detail in Refs.~\cite{Geiger:1991nj,Bass:2002fh},  while production of heavy quarks
has been laid out in Ref.~\cite{Srivastava:2017bcm}. Presently, we shall
%here 
just briefly summarize the features most important to our investigation:

The parton cascade model is a transport model for the time-evolution of an ensemble of quarks and gluons based on the Boltzmann equation.
We include $2\rightarrow 2$ scatterings between light quarks, heavy quarks and gluons, 
and the $2\rightarrow 3$ reactions via time-like branchings of the final-state
partons (see Refs.~\cite{Bass:2002fh,Altarelli:1977zs}) following the well tested procedure adopted in {\tt PYTHIA} \cite{Sjostrand:2006za}.

\begin{figure}
\centerline{\includegraphics*[width=8.0 cm]{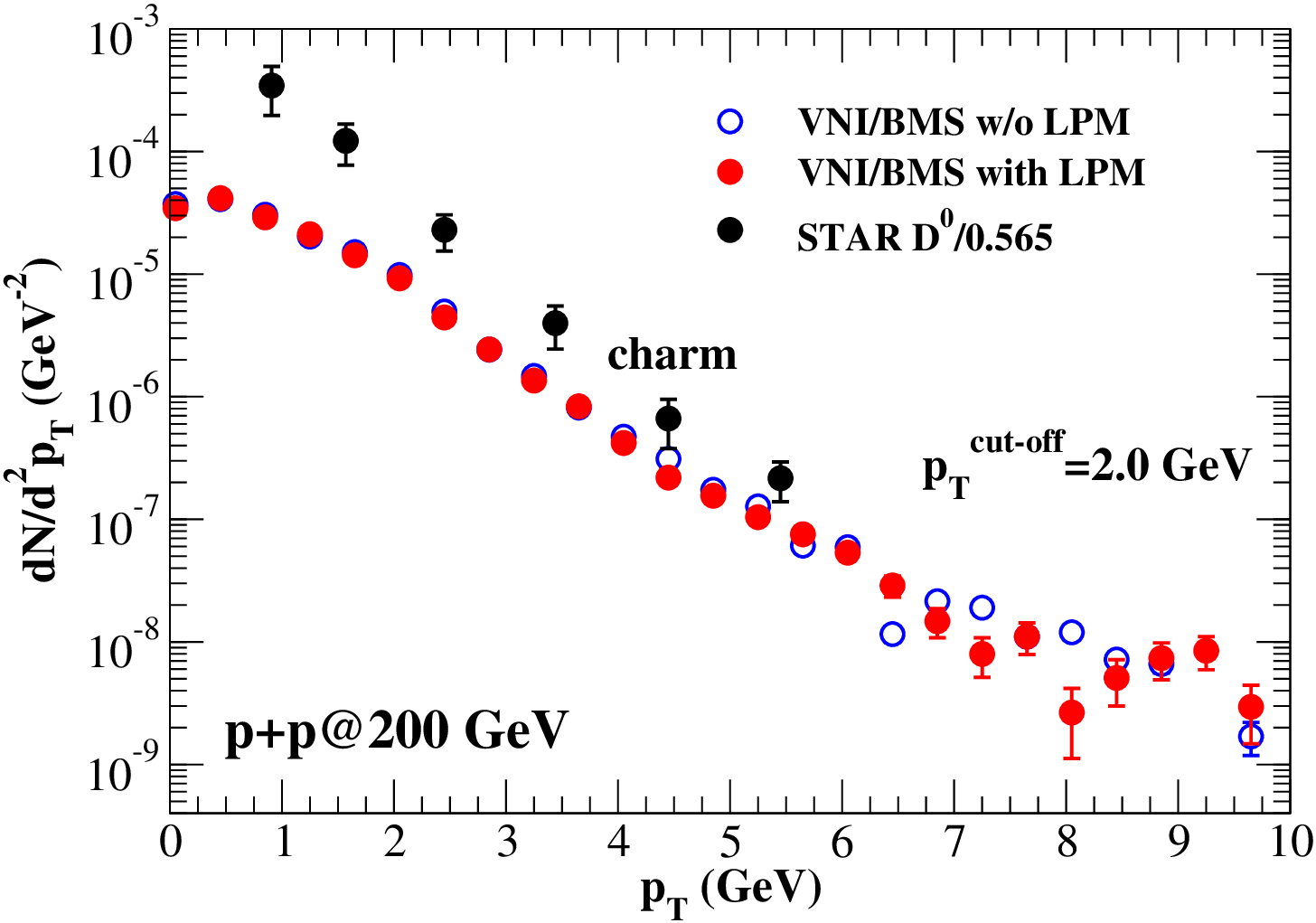}}
\centerline{\includegraphics*[width=8.0 cm]{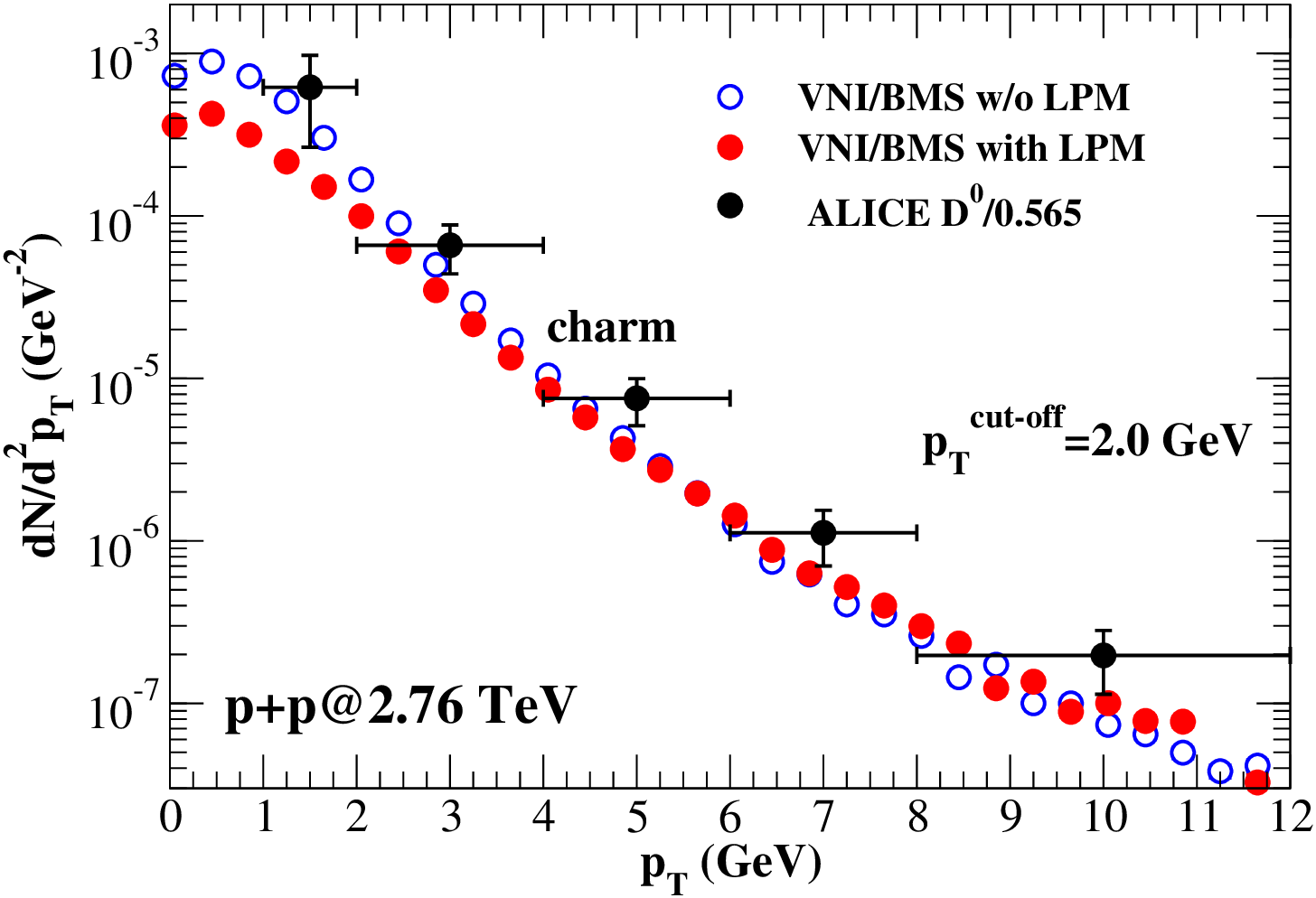}}
\caption{(Color online) The transverse momentum spectra of charm quarks in $pp$ collisions at 
200 GeV (upper panel) and 2.76 TeV (lower panel) due to multiple collisions among partons and fragmentations
off final state quarks, with and without inclusion of LPM effect.}  
\label{0.2_2.76}
\end{figure}

In the PCM, the IR-singularities in these pQCD cross-sections are avoided
by introducing a lower cut-off on the momentum transfer $p_T^\text{cut-off}=$ 2 GeV (please note that results
discussed in Refs.~\cite{Renk:2005yg, Srivastava:2017bcm} were obtained by using a much smaller value
for $p_T^\text{cut-off}$ of about 0.7 GeV, which increased the parton scatterings). 

Most of the studies using {\tt VNI/BMS} reported earlier were
performed using a constant value of $\alpha_s$ = 0.3. In the present work, we have taken 
$\alpha_s(Q^2)$, as  we wish to study the momentum distribution 
of charm quarks for large values of transverse momenta.
Details of the initial parton distributions and the relevant
matrix elements etc. have already been discussed 
repeatedly~\cite{Bass:2002fh,Srivastava:2017bcm,Srivastava:2018dye} which we closely follow.

\begin{figure*}
  \includegraphics[width= 14 cm]{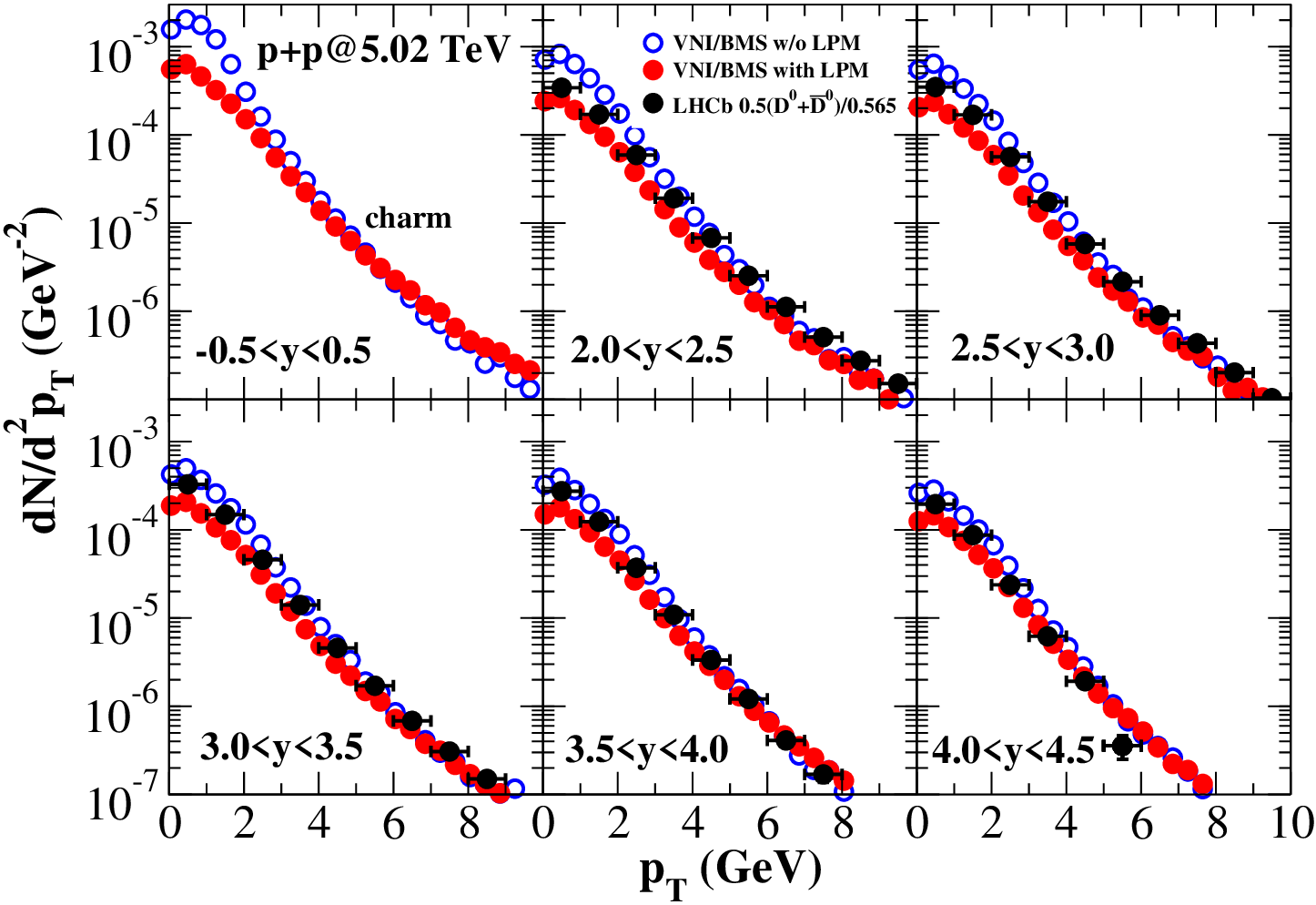}
  \caption{(Color online) The transverse momentum spectra of charm quarks in $pp$ collisions at 
5.02 TeV  due to multiple collisions among partons and fragmentations
off final state quarks, with and without inclusion of LPM effect.}
\label{5.02}
\end{figure*}

For the sake of completeness, we recall that the $2 \rightarrow 3$ processes are 
accounted for by inclusion of radiative processes for the
 final state partons in a leading logarithmic approximation. The collinear singularities 
are then regularized by terminating the time-like branchings, once the virtuality of the parton
drops to $M_0^2(=m_i^2+\mu_0^2)$, where $m_i$ is the current mass of the parton (zero for gluons,
current mass for quarks) and $\mu_0 $ has been kept fixed as 1 GeV. 
We have included $g \rightarrow gg$, $q \rightarrow q g$, 
$g \rightarrow q \bar{q}$, and $q \rightarrow q \gamma$ branchings for 
which the relevant branching functions
$P_{a\rightarrow bc}$ are taken from Altarelli and Parisi~\cite{Altarelli:1977zs}. 
The interference of soft gluons is included by angular ordering of radiated gluons as in 
{\tt PYTHIA}.

Implementing the LPM effect in a semi-classical transport such as the PCM is not easy. First of all, the quarks and gluons are treated
as quasi-particles in the PCM and thus a full quantum mechanical treatment for the process is out of question.
We implement the LPM effect by assigning a formation time $\tau$ to the radiated particle:
\begin{equation}
\tau = \frac{\omega}{k_T^2},
\end{equation}
where $\omega$ is its energy and $k_T$ is its transverse momentum with respect
to the emitter. During the formation time, the radiated particle is assigned zero cross-section
and thus it does not interact. The emitter, however continues to interact and if that happens, the
radiated particle is removed from the list and does not participate in later evolution of the
system. This leads to suppression of parton multiplication 
(see Refs.~\cite{Renk:2005yg,Bass:2002vm,Bass:2007hy}). A similar procedure is 
adopted in the Boltzmann Approach to MultiParton Scattering, {\tt BAMPS},  of the 
Frankfurt group~\cite{Uphoff:2014hza}.
 This particular implementation of the LPM effect is quite common for semi-classical transport models, but by no means unique. An alternative method of implementing the LPM effect by
Baier, Dokshitzer, Mueller, Peigne and Schiff (BDMPS)  relies on recalculating the phase space for the emission of the radiated gluon~\cite{Baier:1996kr,Baier:1996sk,Baier:1998kq} (see also Ref.~\cite{Zakharov:1996fv}). Recently we have experimented with implementing the LPM effect in a scheme that is assured to reproduce the BDMPS limit of parton energy-loss \cite{Wiedemann:2011zz,ColemanSmith:2011wd}. The energy loss
suffered by charm quarks in an infinite medium (at a fixed temperature)
 was well described using this formalism~\cite{Younus:2013rja}. However, this implementation, focussing on the evolution of the leading parton, is currently only feasible for infinite matter calculations in the PCM and further development is required to adapt the necessary algorithms  to proton-proton or nucleus-nucleus calculations.

Our expectation is that the LPM effect will lead to a  suppression of parton multiplication
and thus to a reduction of primary-secondary or seconadary-secondary collisions, where
primary partons make up the initial state of the two colliding protons and secondary partons are the partons
emerging from scatterings and subsequent radiative interactions. It is expected that as the LPM effect reduces the number of multiple scatterings (which mainly produce charm quarks
having low transverse momenta), we should expect
a lowering of the production of charm quarks at smaller $p_T$. In addition, the suppression of radiation of gluons through the LPM effect
should imply that charm quarks having large momenta radiate gluons less frequently. 
This should lead to a hardening
of the transverse momentum spectra for charm quarks. Our analysis is set up to confirm/refute these expectations.

In order to clearly bring out the consequences of the LPM effect we proceed as follows: 
as a first step we study the production of charm quarks with multiple parton collisions and fragmentations  without including the LPM effect. Next we give our results for calculations where the LPM effect is
included.
We investigate whether the LPM effect eliminates multiple parton scatterings by comparing the results from the above to a calculation with only
primary-primary parton scatterings and fragmentations. The difference of the
results of these calculations should clearly bring out the importance
of multiple scatterings of partons in proton-proton collisions and indicate the possible emergence of an interacting medium created by semi-hard pQCD interactions.
 
Finally, in order to investigate the rapidity dependence of the LPM effect,
we shall study the transverse momentum distribution of charm quarks at different rapidities, for which data
have now become available.

\begin{figure*}
  \includegraphics[width= 14 cm]{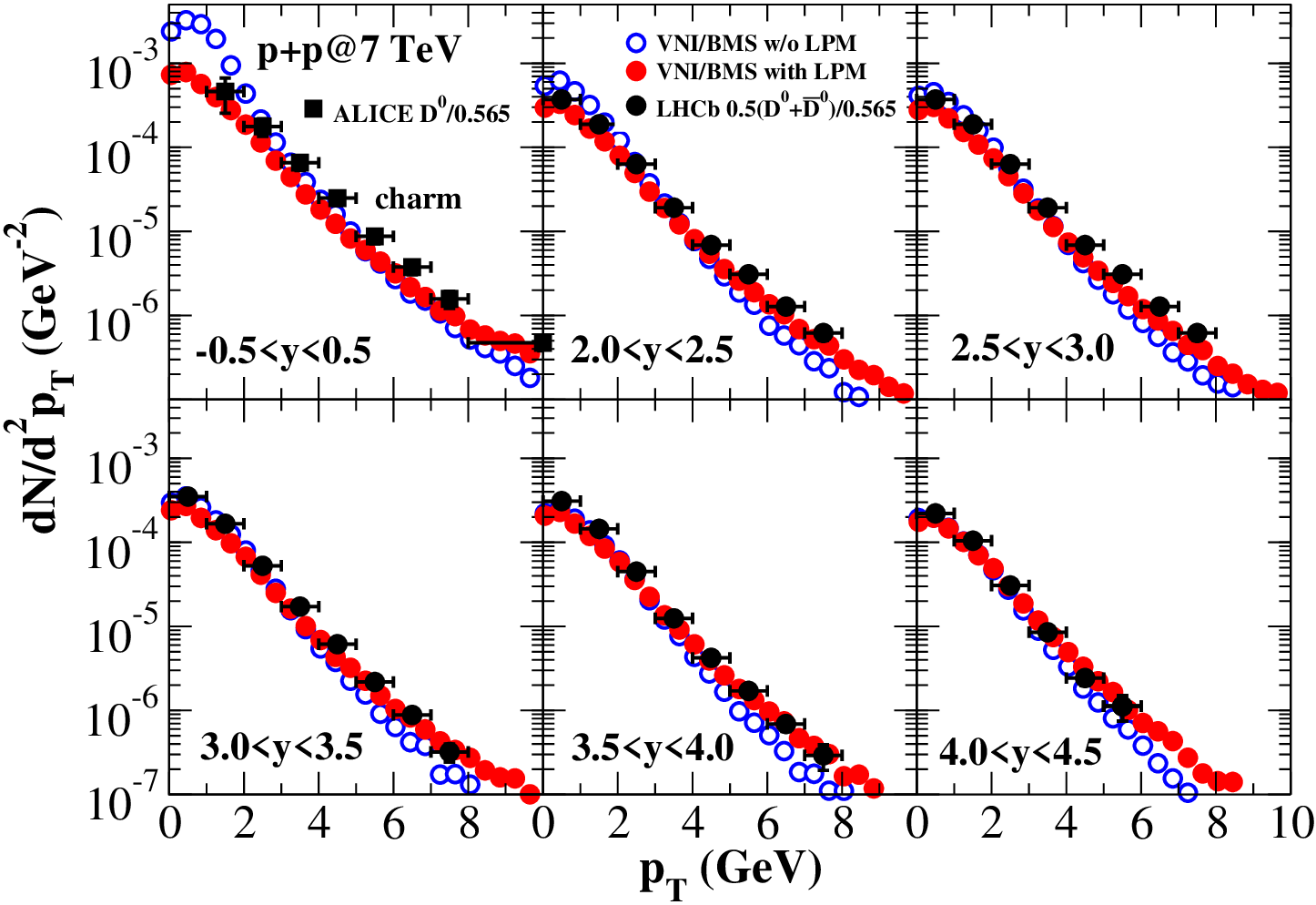}
  \caption{(Color online)The transverse momentum spectra of charm quarks in $pp$ collisions at 
7.00 TeV  due to mutiple collisions among partons and fragmentaions
off final state quarks, with and without inclusion of LPM effect.}
\label{7.00}
\end{figure*}

\section{Results}

\subsection{Multiple scatterings and consequences of LPM effect}

An interacting medium would be characterized by  partons undergoing 
multiple interactions. This is different from the case when we have several parton-parton interactions
involving only primary partons from the projectile and the target, without any further interaction among the
partons thus produced. 
%We can not justifiably call this system as having an interacting medium.

In Fig.~\ref{min-bias} we show results for minimum bias collisions of protons at several
incident beam energies and show the number of semi-hard partonic scatterings, number of fragmentations, and the number of 
charm quarks produced per collision. 

The first set of calculations restricts the interactions to
primary-primary collisions followed by fragmentations off the final state 
partons. These results will not be affected by assigning or not assigning a
formation time (i.e, inclusion or non-inclusion of LPM effect)
 to the radiated gluons as, further scatterings are not considered.
The second set of calculations allows for primary-primary, primary-secondary,
and secondary-secondary collisions along with fragmentations off the 
final state parton, but the LPM effect is not taken into account. 
The final set of calculations describe the system when all possible multiple scatterings and fragmentations
off the final state partons are included and the LPM effect is accounted for, using the procedure discussed
earlier.

We find that without the LPM effect, the number of collisions and fragmentations rise rapidly
with increase in collision energy. The accounting of the LPM effect moderates this rise considerably. 
The reduction in the number of
collisions is about 2\% at 200 GeV and rises to almost 80\% at 13.00 TeV, showing a strong dependence on
the collision energy (for a fixed $p_T^\text{cut-off}$ of 2 GeV). The corresponding reduction in number of
fragmentations is similar, being about 2\% at 200 GeV and rising to about 70\% at 13.00 TeV. The similarity of these
numbers should not come as a surprise as in our approach scatterings are followed by fragmentations. The reduction
in the production of charm quarks is smaller though, just about 1\% at 200 GeV and about 60\% at the top energy considered. 
We attribute the smaller reduction in the charm quark multiplicity compared to the reduction in overall scatterings and fragmentations to the large mass of the charm quark, which restricts the phase space for its production.

As discussed earlier, a comparison between results including the LPM effect with those for only primary-primary
collisions and fragmentations reveal the extent of multiple scatterings. We note that collisions involving 
primary and secondary partons account for about 2\% of the total number of collisions when LPM is accounted for at
200 GeV and increase to about 45\% at the top energy considered. The number of fragmentations also rises similarly.

These results suggest that the semi-hard partonic interactions in $pp$ collisions at
LHC energies produce a dense medium, where partons undergo multiple interactions, even when the LPM
suppression of fragmentations off final state partons is accounted for.
 Theses (additional) multiple collisions are sufficiently large to leave an imprint even in minimum bias events 
which are dominated by collisions involving larger impact parameters where the produced medium may not be very dense.

Evidence of the increasing importance of the LPM effect in more central collisions
(which are likely to have a larger multiplicity) is seen from Fig.~\ref{b.eq.0} where
the corresponding results are plotted for zero impact parameter. We see that the number of collisions, fragmentations,
and charm quarks for all the cases rise significantly and so also the effect of LPM supression. 

\begin{figure*}
  \includegraphics[width= 14 cm]{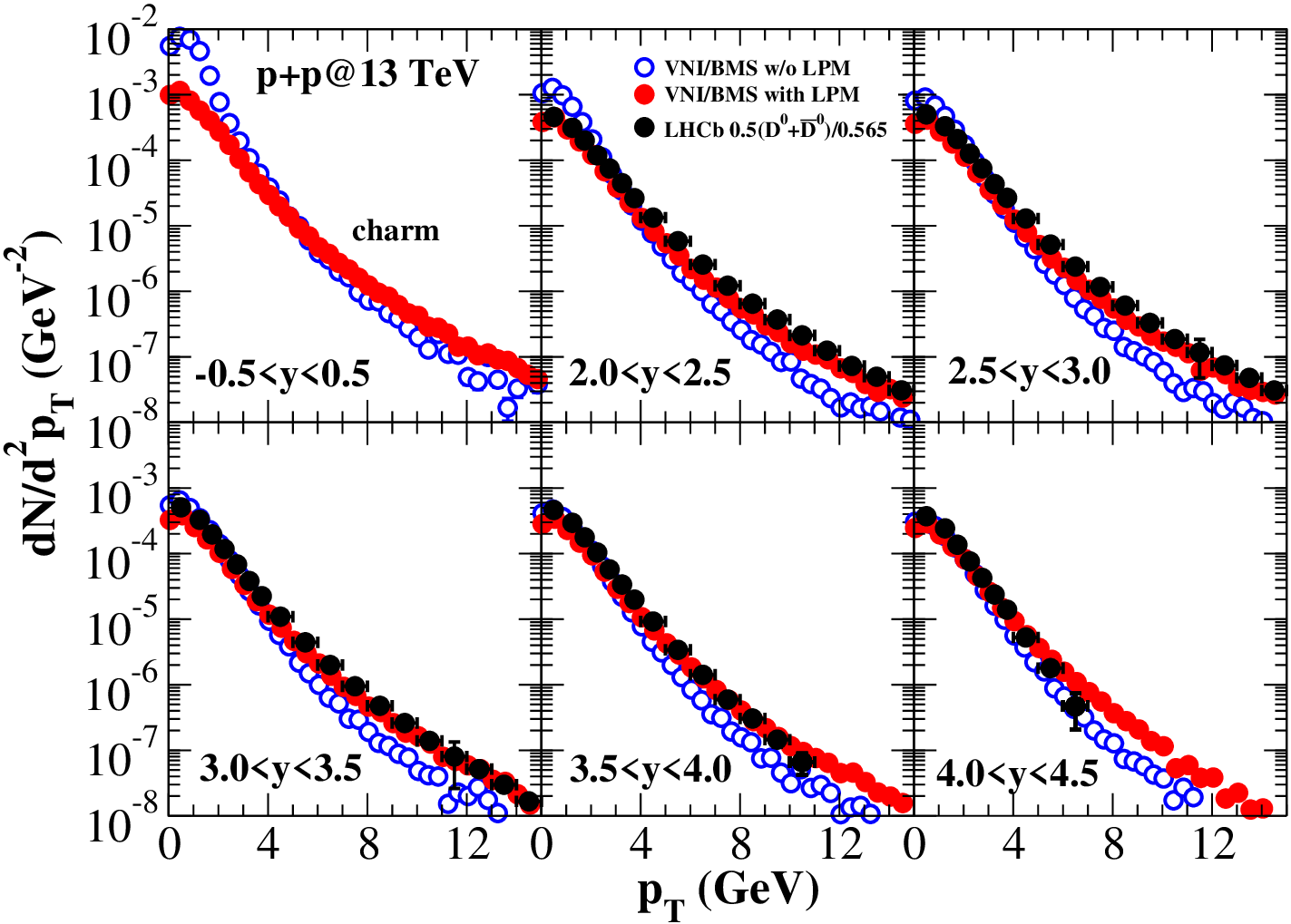}
  \caption{(Color online) The transverse momentum spectra of charm quarks in $pp$ collisions at 
13.00 TeV  due to mutiple collisions among partons and fragmentaions
off final state quarks, with and without inclusion of LPM effect.}
\label{13.00}
\end{figure*}

\subsection{Transverse momentum distribution of charm quarks}

Next we discuss our results for the $p_T$ distribution of charm quarks.
Given the nature of charm quark fragmentation into $D$ mesons, the $p_T$ spectra can be used as a proxy for the $p_T$ distribution of prompt  
$D^0$ mesons, by accounting for the fraction (0.565) for which the charm quark fragments
into a $D^0$ meson. 

We have already seen that the LPM effect has only a very small effect at the lowest incident beam energy considered
here, namely 200 GeV. This is again confirmed by Fig.~\ref{0.2_2.76} (upper panel), where the momentum distribution of
the charm quarks with and without the LPM effect are shown. These are essentially identical.
(The deviation of experimental data~\cite{Adamczyk:2012af} from the theoretical calculations is mainly due to the value of 
$p_T^\text{cut-off}$ of 2 GeV used at all the energies. We believe that a more appropriate value for 
this particular case could be about 0.7 GeV used earlier~\cite{Srivastava:2017bcm}).

The LPM effect starts to become relevant in the theoretical results for the $p_T$ distribution of charm quarks
at 2.76 TeV (see Fig.~\ref{0.2_2.76}, lower panel), where a larger production of charm quarks is 
seen at lower momenta. We note however, that the results
above $p_T$ equal to 2 GeV, where we can trust our results, can not distinguish between the calculations
with and with-out the LPM effect at this beam energy. 
We also add that the agreement of the calculation with the experimental data~\cite{Abelev:2012vra} is 
likely to improve with a slight decrease in $p_T^\text{cut-off}$ 
as it will increase
the number of partonic collisions and the accompanying fragmentations.
 
LHCb has measured charm production at several forward rapidities at 5.02~\cite{Aaij:2016jht}, 
7.00~\cite{Aaij:2013mga}, and 13.00 TeV~\cite{Aaij:2015bpa}. The results at central 
rapidity for the same at 7.00 TeV beam energy from ALICE~\cite{ALICE:2011aa} are also available.

We see a good description of $p_T$ spectra of charm quarks at all rapidities at 5.02 TeV (Fig.~\ref{5.02}). An enhanced
production of charm quarks is seen at lower momenta, when the LPM effect is neglected and the enhancement decreases
with increase in rapidity. It remains to be 
seen if the data likely to be available at 5.02 TeV (and 13.00 TeV, see later) at central rapidity are in agreement
with these results. 

The results for 7.00 TeV (Fig.~\ref{7.00}) are of particular relevance, since experimental data also exist at central rapidity.
Our calculations show a large suppression of charm production at lower $p_T$ when the LPM effect is included and closely reproduce the transverse momentum spectra at all rapidities. We also see a hint of the
hardening of the $p_T$ spectra for large values of $p_T$, which is also reproduced by our calculations,
even though the effect is  not large. As indicated earlier,  this happens as the LPM effect also suppresses the radiation of gluons by charm quarks traversing the medium at large energy/momenta.

The hardening of the transverse momentum spectra and suppression of charm quarks having low $p_T$ (the suppression
decreasing with increase in rapidity) is seen more clearly at 13.00 TeV (Fig.~\ref{13.00}). The experimental results
at all rapidities are adequately explained when the LPM effect is accounted for. It will be interesting to see if the
substantial suppression predicted at central rapidity is supported by data.

\section{Summary and conclusions}

We have studied the impact of the Landau Pomeranchuk Midgal effect on the dynamics of parton transport in proton-proton collisions at LHC energies. In particular, we have focused on the production of charm quarks, since these are only produced in hard pQCD interactions for which the parton cascade model utilized in our study is uniquely suited.
We find that the inclusion of the Landau Pomeranchuk Midgal effect, which suppresses the radiation of gluons off scattered
partons leads to a reduction in the number of scatterings, number of fragmentations and number of charm quarks
which are produced. Even after this suppression, however, these quantities remain larger than the corresponding numbers for calculations 
where only primary-primary collisions among partons is included along with fragmentation off final
state partons.

The results indicate the formation of an interacting medium, which is dense enough for the LPM suppression
of radiation to set in and yet permits multiple scatterings among partons. The LPM effect 
plays an important role in moderating the production of charm quarks having low transverse momenta.
It also leads to a hardening of their transverse momentum 
spectra at larger $p_T$. 
The impact of the LPM effect is found to rise with
increasing collision energy and to decrease with increase in rapidity. 

Before closing, we add that the charm production in $pp$ collisions has been studied in detail using
Fixed Order Next to Leading Log (FONLL) calculations~\cite{Cacciari:1998it}. The data at higher LHC energies 
are generally found to be slightly above  the upper limit given by these calculations 
(see Refs.~\cite{Aaij:2016jht,Aaij:2013mga,Aaij:2015bpa,ALICE:2011aa}). Realizing that
our calculations with only primary-primary collisions and fragmentations tend to roughly
account for the higher order corrections in a Leading Log Approximation, these studies then
suggest additional contributions from multiple scatterings. The precise extent of this contribution
and its dependence on some of the parameters, e.g., current mass of the charm quark, $p_T^\text{cut-off}$
and $\mu_0$ remain to be investigated. We do believe, however, that the additional contributions
arising due to the multiple scatterings and suject to LPM effect will be there, unless of-course
$p_T^\text{cut-off}$ and $\mu_0$ are taken too large and too few interactions take place and
too few radiations occur.

In brief, our results provide an indication of emergence of a dense and interacting medium
of partons in $pp$ collisions at LHC energies due to semi-hard pQCD interactions, even when 
LPM suppression of radiation of gluons from scattered partons is accounted for.

\section*{Acknowledgments} 
DKS gratefully acknowledges the support by the Department of Atomic
Energy. This research was supported in part by the ExtreMe Matter Institute EMMI at the GSI Helmholtzzentrum für Schwerionenforschung, Darmstadt,
Germany. DKS also acknowledges valuable discussions with Jurgen Schukraft and Dirk Rischke.
SAB acknowledges support by US Department of Energy grant DE-FG02-05ER41367.

%\newpage

%\bibliographystyle{h-physrev5}
%\bibliography{/Users/bass/Github/Bibliography/Duke_QCD_refs}

\begin{thebibliography}{10}

%\cite{Ratti:2016lrh}
\bibitem{Ratti:2016lrh}
  C.~Ratti,
  %``Recent results on lattice QCD thermodynamics,''
  \newblock J.\ Phys.\ Conf.\ Ser.\  {\bf 736}, no.1,  012001 (2016).
  doi:10.1088/1742-6596/736/1/012001
  %%CITATION = doi:10.1088/1742-6596/736/1/012001;%%

%\cite{Ratti:2017qgq}
\bibitem{Ratti:2017qgq}
  C.~Ratti,
  %``Bulk properties of QCD matter from lattice simulations,''
  \newblock J.\ Phys.\ Conf.\ Ser.\  {\bf 779}, no.1,  012016 (2017).
  doi:10.1088/1742-6596/779/1/012016
  %%CITATION = doi:10.1088/1742-6596/779/1/012016;%%

%\cite{Ratti:2018ksb}
\bibitem{Ratti:2018ksb}
  C.~Ratti,
  %``Lattice QCD and heavy ion collisions: a review of recent progress,''
  arXiv:1804.07810 [hep-lat].
  %%CITATION = ARXIV:1804.07810;%%

\bibitem{Schenke:2010rr}
  B.~Schenke, S.~Jeon and C.~Gale,
  %``Elliptic and triangular flow in event-by-event (3+1)D viscous hydrodynamics,''
 \newblock Phys.\ Rev.\ Lett.\  {\bf 106}, 042301 (2011)
  doi:10.1103/PhysRevLett.106.042301
  [arXiv:1009.3244 [hep-ph]].
  %%CITATION = doi:10.1103/PhysRevLett.106.042301;%%

\bibitem{Gale:2012rq}
  C.~Gale, S.~Jeon, B.~Schenke, P.~Tribedy and R.~Venugopalan,
  %``Event-by-event anisotropic flow in heavy-ion collisions from combined Yang-Mills and viscous fluid dynamics,''
 \newblock Phys.\ Rev.\ Lett.\  {\bf 110}, no.1,  012302 (2013)
  doi:10.1103/PhysRevLett.110.012302
  [arXiv:1209.6330 [nucl-th]].
  %%CITATION = doi:10.1103/PhysRevLett.110.012302;%%

%\cite{Shen:2014vra}
\bibitem{Shen:2014vra}
  C.~Shen, Z.~Qiu, H.~Song, J.~Bernhard, S.~Bass and U.~Heinz,
  %``The iEBE-VISHNU code package for relativistic heavy-ion collisions,''
 \newblock Comput.\ Phys.\ Commun.\  {\bf 199}, 61 (2016)
  doi:10.1016/j.cpc.2015.08.039
  [arXiv:1409.8164 [nucl-th]].
  %%CITATION = doi:10.1016/j.cpc.2015.08.039;%%

%\cite{Bernhard:2016tnd}
\bibitem{Bernhard:2016tnd}
  J.~E.~Bernhard, J.~S.~Moreland, S.~A.~Bass, J.~Liu and U.~Heinz,
  %``Applying Bayesian parameter estimation to relativistic heavy-ion collisions: simultaneous characterization of the initial state and quark-gluon plasma medium,''
 \newblock Phys.\ Rev.\ C {\bf 94}, no.2,  024907 (2016)
  doi:10.1103/PhysRevC.94.024907
  [arXiv:1605.03954 [nucl-th]].
  %%CITATION = doi:10.1103/PhysRevC.94.024907;%%

%\cite{Khachatryan:2016txc}
\bibitem{Khachatryan:2016txc}
  V.~Khachatryan {\it et al.} [CMS Collaboration],
  %``Evidence for collectivity in pp collisions at the LHC,''
 \newblock Phys.\ Lett. {\bf B765},  193 (2017)
  doi:10.1016/j.physletb.2016.12.009
  [arXiv:1606.06198 [nucl-ex]].
  %%CITATION = doi:10.1016/j.physletb.2016.12.009;%%

%\cite{ALICE:2017jyt}
\bibitem{ALICE:2017jyt}
  J.~Adam {\it et al.} [ALICE Collaboration],
  %``Enhanced production of multi-strange hadrons in high-multiplicity proton-proton collisions,''
  \newblock Nature Phys.\  {\bf 13}, 535 (2017)
  doi:10.1038/nphys4111
  [arXiv:1606.07424 [nucl-ex]].
  %%CITATION = doi:10.1038/nphys4111;%%

%\cite{Srivastava:2018dye}
\bibitem{Srivastava:2018dye}
  D.~K.~Srivastava, R.~Chatterjee and S.~A.~Bass,
  %``Transport Dynamics of Parton Interactions in pp Collisions at LHC Energies,''
  \newblock Phys.\ Rev.\ C {\bf 97},  no.6,  064910 (2018)
  doi:10.1103/PhysRevC.97.064910
  [arXiv:1801.07482 [nucl-th]].

\bibitem{Geiger:1991nj}
K.~Geiger and B.~Muller,
\newblock Nucl. Phys. {\bf B369}, 600 (1992).
%%CITATION = NUPHA,B369,600;%%

\bibitem{Bass:2002fh}
S.~A. Bass, B.~Muller, and D.~K. Srivastava,
\newblock Phys. Lett. {\bf B551}, 277 (2003), nucl-th/0207042.
%%CITATION = NUCL-TH 0207042;%%

\bibitem{Altarelli:1977zs}
G.~Altarelli and G.~Parisi,
\newblock Nucl. Phys. {\bf B126}, 298 (1977).
%%CITATION = NUPHA,B126,298;%%

\bibitem{Landau:1953gr}
  L.~D.~Landau and I.~Pomeranchuk,
  \newblock Dokl.\ Akad.\ Nauk Ser.\ Fiz.\  {\bf 92}, 735 (1953),
    L.~D.~Landau and I.~Pomeranchuk,
  \newblock Dokl.\ Akad.\ Nauk Ser.\ Fiz.\  {\bf 92}, 535 (1953),
   A.~B.~Midgal, 
  \newblock Phys. Rev. {\bf 103}, 1811 (1956).

\bibitem{Srivastava:2017bcm}
  D.~K.~Srivastava, S.~A.~Bass and R.~Chatterjee,
  %``Production of charm quarks in a parton cascade model for relativistic heavy 
  % ion collisions at $\sqrt{s_{\textrm NN}}$= 200 GeV,''
  \newblock Phys.\ Rev.\ C {\bf 96}, 064906 (2017), 
  doi:10.1103/PhysRevC.96.064906, 
  [arXiv:1705.05542 [nucl-th]].
%\cite{Srivastava:2018dye}

%\cite{Bethe:1934za}
\bibitem{Bethe:1934za}
  H.~Bethe and W.~Heitler,
  %``On the Stopping of fast particles and on the creation of positive electrons,''
  \newblock Proc.\ Roy.\ Soc.\ Lond.\ A {\bf 146}, 83 (1934).
  doi:10.1098/rspa.1934.0140
  %%CITATION = doi:10.1098/rspa.1934.0140;%%


\bibitem{Renk:2005yg}
T.~Renk, S.~A. Bass, and D.~K.~Srivastava,
\newblock Phys. Lett. {\bf B632}, 632 (2006), nucl-th/0505059.
%%CITATION = NUCL-TH 0505059;%%

\bibitem{Bass:2002vm}
S.~A. Bass, B.~Muller, and D.~K.~Srivastava,
\newblock Phys. Rev. Lett. {\bf 91}, 052302 (2003), nucl-th/0212103.
%%CITATION = NUCL-TH 0212103;%%

%\cite{%\cite{Bass:2007hy}
\bibitem{Bass:2007hy}
  S.~A.~Bass, T.~Renk and D.~K.~Srivastava,
  %``Photon production in the parton cascade model,''
 \newblock Nucl.\ Phys. {\bf A783}, 367 (2007).
  doi:10.1016/j.nuclphysa.2006.11.029
  %%CITATION = doi:10.1016/j.nuclphysa.2006.11.029;%%
 

\bibitem{Sjostrand:2006za}
  T.~Sjostrand, S.~Mrenna and P.~Z.~Skands,
  %``PYTHIA 6.4 Physics and Manual,''
 \newblock JHEP {\bf 0605}, 026 (2006)
  doi:10.1088/1126-6708/2006/05/026
  [hep-ph/0603175].
  %%CITATION = doi:10.1088/1126-6708/2006/05/026;%%

%\cite{Uphoff:2014hza}
\bibitem{Uphoff:2014hza} 
  J.~Uphoff, O.~Fochler, Z.~Xu and C.~Greiner,
  %``Elastic and radiative heavy quark interactions in ultra-relativistic heavy-ion collisions,''
  \newblock J.\ Phys.\ G {\bf 42}, no. 11, 115106 (2015)
  doi:10.1088/0954-3899/42/11/115106
  [arXiv:1408.2964 [hep-ph]].
  %%CITATION = doi:10.1088/0954-3899/42/11/115106;%%

\bibitem{Baier:1996kr}
  R.~Baier, Y.~L.~Dokshitzer, A.~H.~Mueller, S.~Peigne and D.~Schiff,
  %``Radiative energy loss of high-energy quarks and gluons in a finite volume quark - gluon plasma,''
  \newblock Nucl.\ Phys. {\bf B483}, 291  (1997)
  doi:10.1016/S0550-3213(96)00553-6
  [hep-ph/9607355].
  %%CITATION = doi:10.1016/S0550-3213(96)00553-6;%%

%\cite{Baier:1996sk}
\bibitem{Baier:1996sk}
  R.~Baier, Y.~L.~Dokshitzer, A.~H.~Mueller, S.~Peigne and D.~Schiff,
  %``Radiative energy loss and p(T) broadening of high-energy partons in nuclei,''
  Nucl.\ Phys. {\bf B484}, 265 (1997) 
  doi:10.1016/S0550-3213(96)00581-0
  [hep-ph/9608322].
  %%CITATION = doi:10.1016/S0550-3213(96)00581-0;%%

%\cite{Baier:1998kq}
\bibitem{Baier:1998kq}
  R.~Baier, Y.~L.~Dokshitzer, A.~H.~Mueller and D.~Schiff,
  %``Medium induced radiative energy loss: Equivalence between the BDMPS and Zakharov formalisms,''
  Nucl.\ Phys. {\bf B531}, 403  (1998)
  doi:10.1016/S0550-3213(98)00546-X
  [hep-ph/9804212].
  %%CITATION = doi:10.1016/S0550-3213(98)00546-X;%%

%\cite{Zakharov:1996fv}
\bibitem{Zakharov:1996fv}
  B.~G.~Zakharov,
  %``Fully quantum treatment of the Landau-Pomeranchuk-Migdal effect in QED and QCD,''
  JETP Lett.\  {\bf 63}, 952  (1996)
  doi:10.1134/1.567126
  [hep-ph/9607440].
  %%CITATION = doi:10.1134/1.567126;%%

%\cite{Wiedemann:2011zz}
\bibitem{Wiedemann:2011zz}
  U.~A.~Wiedemann, K.~C.~Zapp and J.~Stachel,
  %``A Monte Carlo implementation of the BDMPS-Z formalism,''
  \newblock Nucl.\ Phys. {\bf A855}, 285 (2011).
  doi:10.1016/j.nuclphysa.2011.02.060
  %%CITATION = doi:10.1016/j.nuclphysa.2011.02.060;%%

%\cite{ColemanSmith:2011wd}
\bibitem{ColemanSmith:2011wd}
  C.~E.~Coleman-Smith, S.~A.~Bass and D.~K.~Srivastava,
  %``Implementing the LPM effect in a parton cascade model,''
  \newblock Nucl.\ Phys. {\bf A862-863}, 275  (2011)
  doi:10.1016/j.nuclphysa.2011.05.071
  [arXiv:1101.4895 [hep-ph]].
  %%CITATION = doi:10.1016/j.nuclphysa.2011.05.071;%%

%\cite{Younus:2013rja}
\bibitem{Younus:2013rja} 
  M.~Younus, C.~E.~Coleman-Smith, S.~A.~Bass and D.~K.~Srivastava,
  %``Charm Quark Energy Loss In Infinite QCD Matter Using A Parton Cascade Model,''
  \newblock Phys.\ Rev.\ C {\bf 91}, 024912 (2015)
  doi:10.1103/PhysRevC.91.024912
  [arXiv:1309.1276 [nucl-th]].
  %%CITATION = doi:10.1103/PhysRevC.91.024912;%%


%\cite{Adamczyk:2012af}
\bibitem{Adamczyk:2012af}
  L.~Adamczyk {\it et al.} [STAR Collaboration],
  %``Measurements of $D^{0}$ and $D^{*}$ Production in $p+p$ Collisions at $\sqrt{s} = 200$ GeV,''
  \newblock Phys.\ Rev.\ D {\bf 86},  072013 (2012),
  doi:10.1103/PhysRevD.86.072013
  [arXiv:1204.4244 [nucl-ex]].
  %%CITATION = doi:10.1103/PhysRevD.86.072013;%%
  %113 citations counted in INSPIRE as of 04 Apr 2018

\bibitem{Abelev:2012vra}
  B.~Abelev {\it et al.} [ALICE Collaboration],
  %``Measurement of charm production at central rapidity in proton-proton collisions at $\sqrt{s}=2.76$ TeV,''
\newblock  JHEP {\bf 1207}, 191 (2012),
  doi:10.1007/JHEP07(2012)191
  [arXiv:1205.4007 [hep-ex]].

\bibitem{Aaij:2016jht}
  R.~Aaij {\it et al.} [LHCb Collaboration],
  %``Measurements of prompt charm production cross-sections in pp collisions at $ \sqrt{s}=5 $ TeV,''
  \newblock JHEP {\bf 1706},  147 (2017),
  doi:10.1007/JHEP06(2017)147
  [arXiv:1610.02230 [hep-ex]].
  %%CITATION = doi:10.1007/JHEP06(2017)147;%%
  %27 citations counted in INSPIRE as of 04 Apr 2018


%\cite{Aaij:2013mga}
\bibitem{Aaij:2013mga}
  R.~Aaij {\it et al.} [LHCb Collaboration],
  %``Prompt charm production in pp collisions at sqrt(s)=7 TeV,''
  \newblock Nucl.\ Phys. {\bf B871},  1 (2013)
  doi:10.1016/j.nuclphysb.2013.02.010
  [arXiv:1302.2864 [hep-ex]].
  %%CITATION = doi:10.1016/j.nuclphysb.2013.02.010;%%
  %224 citations counted in INSPIRE as of 04 Apr 2018


%\cite{Aaij:2015bpa}
\bibitem{Aaij:2015bpa}
  R.~Aaij {\it et al.} [LHCb Collaboration],
  %``Measurements of prompt charm production cross-sections in $pp$ collisions at $ \sqrt{s}=13 $ TeV,''
  \newblock JHEP {\bf 1603},  159 (2016);
   Erratum: [JHEP {\bf 1609}, 013 (2016)] ;
   Erratum: [JHEP {\bf 1705}, 074 (2017)],
  doi:10.1007/JHEP03(2016)159, 10.1007/JHEP09(2016)013, 10.1007/JHEP05(2017)074
  [arXiv:1510.01707 [hep-ex]].
  %%CITATION = doi:10.1007/JHEP03(2016)159, 10.1007/JHEP09(2016)013, 10.1007/JHEP05(2017)074;%%
  %108 citations counted in INSPIRE as of 04 Apr 2018

\bibitem{ALICE:2011aa}
B.~Abelev {\it et al.} [ALICE Collaboration],
  %``Measurement of charm production at central rapidity in proton-proton collisions at $\sqrt{s} = 7$ TeV,''
\newblock  JHEP {\bf 1201}, 128 (2012),
  doi:10.1007/JHEP01(2012)128 
  [arXiv:1111.1553 [hep-ex]]

%\cite{Cacciari:1998it}
\bibitem{Cacciari:1998it}
  M.~Cacciari, M.~Greco and P.~Nason,
  %``The P(T) spectrum in heavy flavor hadroproduction,''
  JHEP {\bf 9805},  007 (1998)
  doi:10.1088/1126-6708/1998/05/007
  [hep-ph/9803400].
  %%CITATION = doi:10.1088/1126-6708/1998/05/007;%%




\end{thebibliography}

\end{document}